\documentclass[prl,twocolumn,superscriptaddress,showpacs,preprintnumbers]{revtex4-1}
\usepackage{amsmath,bm,graphicx,hyperref}

\renewcommand\d{\partial}
\newcommand\grad{\bm{\nabla}}
\newcommand\+{\dagger}
\renewcommand\k{{\bm{k}}}
\newcommand\p{{\bm{p}}}
\newcommand\q{{\bm{q}}}
\newcommand\sect[1]{\subsection{#1}}

\begin{document}
\preprint{LA-UR-13-20038, NT@UW-13-01, EFI-13-01}

\title{Super Efimov effect of resonantly interacting fermions in two dimensions}

\author{Yusuke Nishida}
\affiliation{Theoretical Division, Los Alamos National Laboratory,
Los Alamos, New Mexico 87545, USA}
\author{Sergej Moroz}
\affiliation{Department of Physics, University of Washington,
Seattle, Washington 98195, USA}
\author{Dam Thanh Son}
\affiliation{Enrico Fermi Institute, University of Chicago,
Chicago, Illinois 60637, USA}

\date{January 2013}

\begin{abstract}
 We study a system of spinless fermions in two dimensions with a
 short-range interaction fine-tuned to a $p$-wave resonance.  We show
 that three such fermions form an infinite tower of bound states of
 orbital angular momentum $\ell=\pm1$ and their binding energies obey a
 universal doubly exponential scaling
 $E_3^{(n)}\propto\exp\bigl(-2e^{3\pi n/4+\theta}\bigr)$ at large $n$.
 This ``super Efimov effect'' is found by a renormalization group
 analysis and confirmed by solving the bound state problem.  We also
 provide an indication that there are $\ell=\pm2$ four-body resonances
 associated with every three-body bound state at
 $E_4^{(n)}\propto\exp\bigl(-2e^{3\pi n/4+\theta-0.188}\bigr)$.  These
 universal few-body states may be observed in ultracold atom experiments
 and should be taken into account in future many-body studies of the
 system.
\end{abstract}


\maketitle

\sect{Introduction}
Recently topological superconductors have attracted great interest
across many subfields in physics~\cite{Hasan:2010,Qi:2011}.  This is
partially because vortices in topological superconductors bind
zero-energy Majorana fermions and obey non-Abelian statistics, which can
be of potential use for fault-tolerance topological quantum
computation~\cite{Kitaev:2003,Nayak:2008}.  A canonical example of such
topological superconductors is a $p$-wave paired state of spinless
fermions in two dimensions~\cite{Read:2002}, which is believed to be
realized in Sr$_2$RuO$_4$~\cite{Mackenzie:2003}.  Previous mean-field
studies revealed that a topological quantum phase transition takes place
across a $p$-wave Feshbach
resonance~\cite{Gurarie:2005,Botelho:2005,Cheng:2005}.

\begin{table}[b]
 \caption{Comparison of the Efimov effect versus the super Efimov
 effect.  \label{tab:comparison}}
 \begin{ruledtabular}
  \begin{tabular}{ccccc}
   & Efimov effect && Super Efimov effect & \\[2pt]\hline
   & Three bosons && Three fermions & \\
   & Three dimensions && Two dimensions & \\
   & $s$-wave resonance && $p$-wave resonance & \\
   & $\ell=0$ && $\ell=\pm1$ & \\
   & Exponential scaling && Doubly exponential scaling &
  \end{tabular}
 \end{ruledtabular}
\end{table}

In this Letter, we study few-body physics of spinless fermions in two
dimensions right at the $p$-wave resonance.  We predict that three such
fermions form an infinite tower of bound states of orbital angular
momentum $\ell=\pm1$ and their binding energies obey a universal doubly
exponential scaling
\begin{align}\label{eq:3-body_energy}
 E_3^{(n)} \propto \exp\bigl(-2e^{3\pi n/4+\theta}\bigr)
\end{align}
at large $n$.  Here $\theta$ is a nonuniversal constant defined modulo
$3\pi/4$.  This novel phenomenon shall be termed a super Efimov effect,
because it resembles the Efimov effect in which three spinless bosons in
three dimensions right at an $s$-wave resonance form an infinite tower
of $\ell=0$ bound states whose binding energies obey the universal
exponential scaling $E_3^{(n)} \propto e^{-2\pi n/s_0}$ with
$s_0\approx1.00624$~\cite{Efimov:1970} (see Table~\ref{tab:comparison}
for comparison).  While the Efimov effect is possible in other
situations~\cite{Efimov:1973,Nishida:2011}, it does not take place in
two dimensions or with $p$-wave
interactions~\cite{Lim:1980,Nishida:2011,Nishida:2012}.  We also provide
an indication that there are $\ell=\pm2$ four-body resonances associated
with every three-body bound state at
\begin{align}\label{eq:4-body_energy}
 E_4^{(n)} \propto \exp\bigl(-2e^{3\pi n/4+\theta-0.188}\bigr),
\end{align}
which also resembles the pair of four-body resonances in the usual
Efimov effect~\cite{Hammer:2007,von-Stecher:2009}.  These universal
few-body states of resonantly interacting fermions in two dimensions
should be taken into account in future many-body studies beyond the
mean-field approximation.

\sect{Renormalization group analysis}
The above predictions can be derived most conveniently by a
renormalization group (RG) analysis.  The most general Lagrangian
density that includes up to marginal couplings consistent with rotation
and parity symmetries is
\begin{align}\label{eq:lagrangian}
 \mathcal{L} &= \psi^\+\!\left(i\d_t+\frac{\grad^2}2\right)\!\psi
 + \phi_a^\+\!\left(i\d_t+\frac{\grad^2}4-\varepsilon_0\right)\!\phi_a \notag\\[3pt]
 & + g\,\phi_a^\+\psi(-i\nabla_a)\psi
 + g\,\psi^\+(-i\nabla_{-a})\psi^\+\phi_a \notag\\[4pt]
 & + v_3\,\psi^\+\phi_a^\+\phi_a\psi + v_4\,\phi_a^\+\phi_{-a}^\+\phi_{-a}\phi_a
 + v_4'\,\phi_a^\+\phi_a^\+\phi_a\phi_a.
\end{align}
Here and below, $\hbar=m=1$, $\nabla_\pm\equiv\nabla_x\pm i\nabla_y$,
and sums over repeated indices $a=\pm$ are assumed.  $\psi$ and
$\phi_\pm$ fields correspond to a spinless fermion and $\ell=\pm1$
composite boson, respectively.  The $p$-wave resonance is defined by the
divergence of the two-fermion scattering amplitude at zero energy, which
is achieved by tuning the bare detuning parameter at
$\varepsilon_0=g^2\Lambda^2/(2\pi)$ with $\Lambda$ being a momentum
cutoff.

\begin{figure}[t]
 \includegraphics[width=0.5\columnwidth,clip]{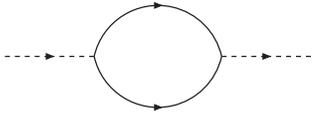}
 \caption{Self-energy diagram of the $\phi_a$ field.  Solid and dashed
 lines represent propagators of $\psi$ and $\phi_a$ fields,
 respectively.  \label{fig:2-body}}
\end{figure}

Since the other couplings ($g$, $v_3$, $v_4$, $v_4'$) are all
dimensionless, our effective field theory (\ref{eq:lagrangian}) is
renormalizable, and its renormalization can be performed in a similar
way to Ref.~\cite{Nishida:2008}.  The self-energy diagram of the
$\phi_a$ field depicted in Fig.~\ref{fig:2-body} is logarithmically
divergent and renormalizes the two-body coupling $g$.  The RG equation
that governs the running of $g$ at a momentum scale $e^{-s}\Lambda$ is
found to be
\begin{align}
 \frac{d g}{d s} = -\frac{g^3}{2\pi},
\end{align}
which is solved by
\begin{align}
 g^2(s) = \frac1{\frac{s}\pi+\frac1{g^2(0)}}.
\end{align}
Accordingly, the two-body coupling is marginally irrelevant; i.e., it
gets weak toward the infrared limit $s\to\infty$.

\begin{figure}[b]
 \includegraphics[width=0.7\columnwidth,clip]{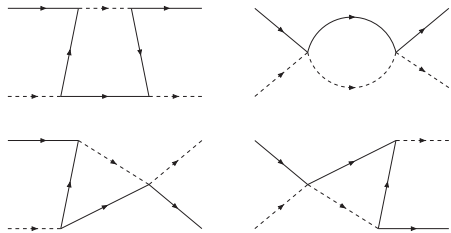}
 \caption{Diagrams to renormalize the three-body coupling $v_3$.
 \label{fig:3-body}}
\end{figure}

We now turn to the renormalization of the three-body coupling $v_3$.
Diagrams that renormalize $v_3$ are depicted in Fig.~\ref{fig:3-body}.
By taking into account the contribution from the $\phi_a$ field
renormalization, the RG equation that governs the running of $v_3$ is
found to be
\begin{align}\label{eq:3-body_RG}
 \frac{dv_3}{ds} = \frac{16g^4}{3\pi} - \frac{11g^2v_3}{3\pi} + \frac{2v_3^2}{3\pi}.
\end{align}
In the infrared limit $s\to\infty$, where the two-body coupling scales as
$g^2\to\pi/s$~\cite{Moroz:2009}, the analytic solution to
Eq.~(\ref{eq:3-body_RG}) becomes
\begin{align}
 v_3(s) \to \frac{2\pi}s\left[1-\cot\!\left(\frac43(\ln s-\theta)\right)\right],
\end{align}
where the angle $\theta$ depends on the initial conditions of $g$ and
$v_3$ at the ultraviolet scale $s\sim0$.

We find that $sv_3$ at large $s$ is a periodic function of $\ln s$ and
diverges at $\ln s=3\pi n/4+\theta$.  These divergences in the
three-body coupling indicate the existence of an infinite tower of
energy scales in the three-body system in the $\ell=\pm1$ channels.
Following the usual Efimov effect~\cite{Bedaque:1999}, we identify these
scales with binding energies which leads to the prediction presented in
Eq.~(\ref{eq:3-body_energy}).  This identification will be confirmed
later by solving the bound state problem.

\begin{figure}[t]
 \includegraphics[width=\columnwidth,clip]{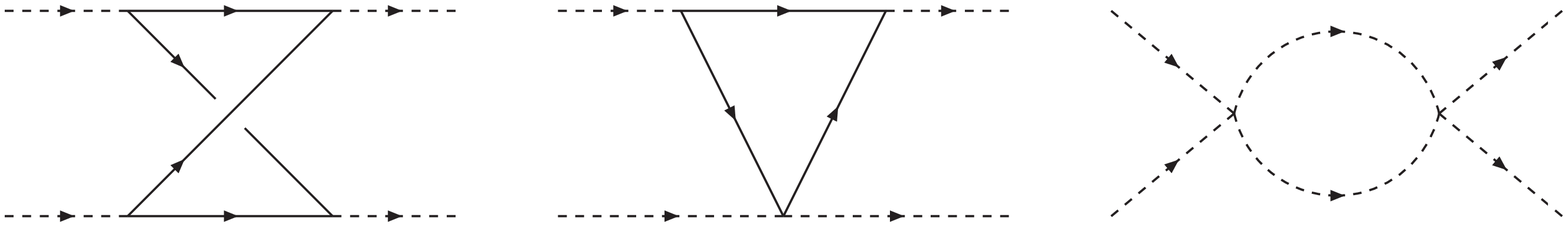}
 \caption{Diagrams to renormalize the four-body couplings $v_4$ and $v_4'$.
 \label{fig:4-body}}
\end{figure}

\begin{figure}[b]
 \includegraphics[width=0.8\columnwidth,clip]{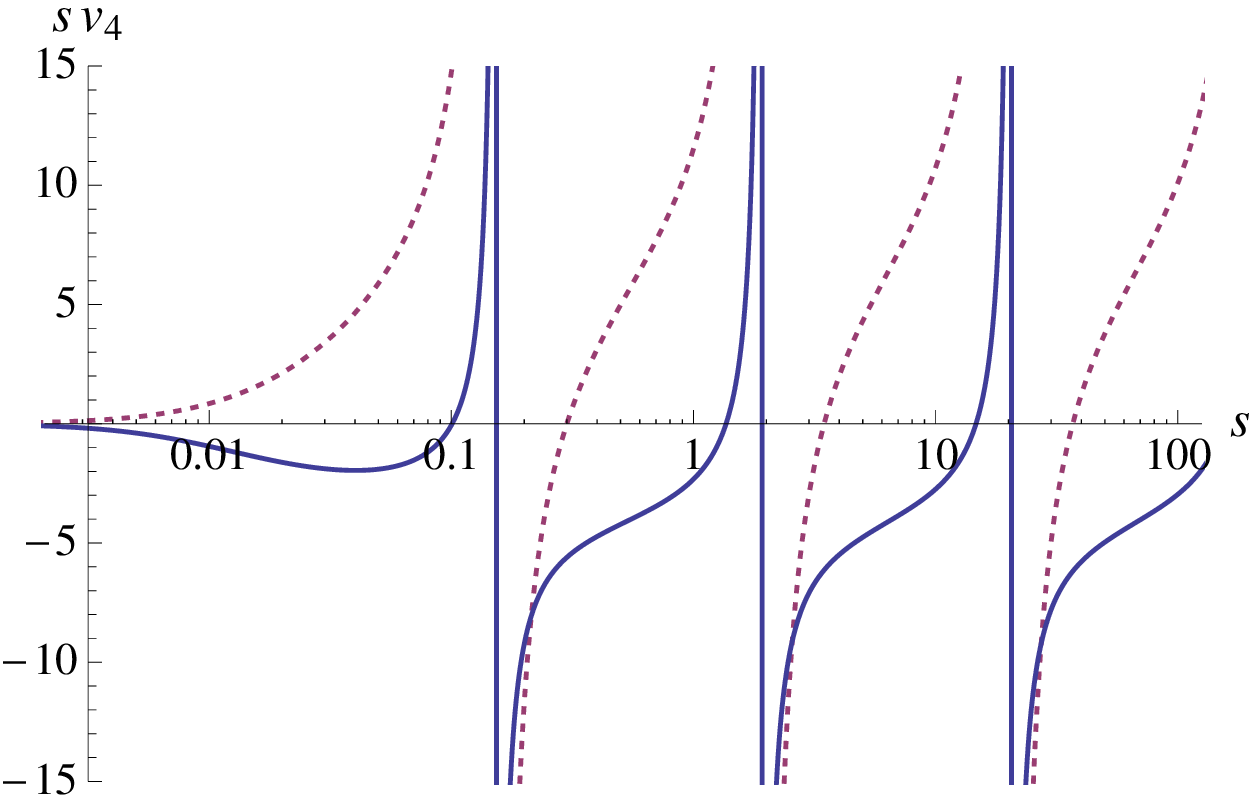}\medskip\\
 \includegraphics[width=0.8\columnwidth,clip]{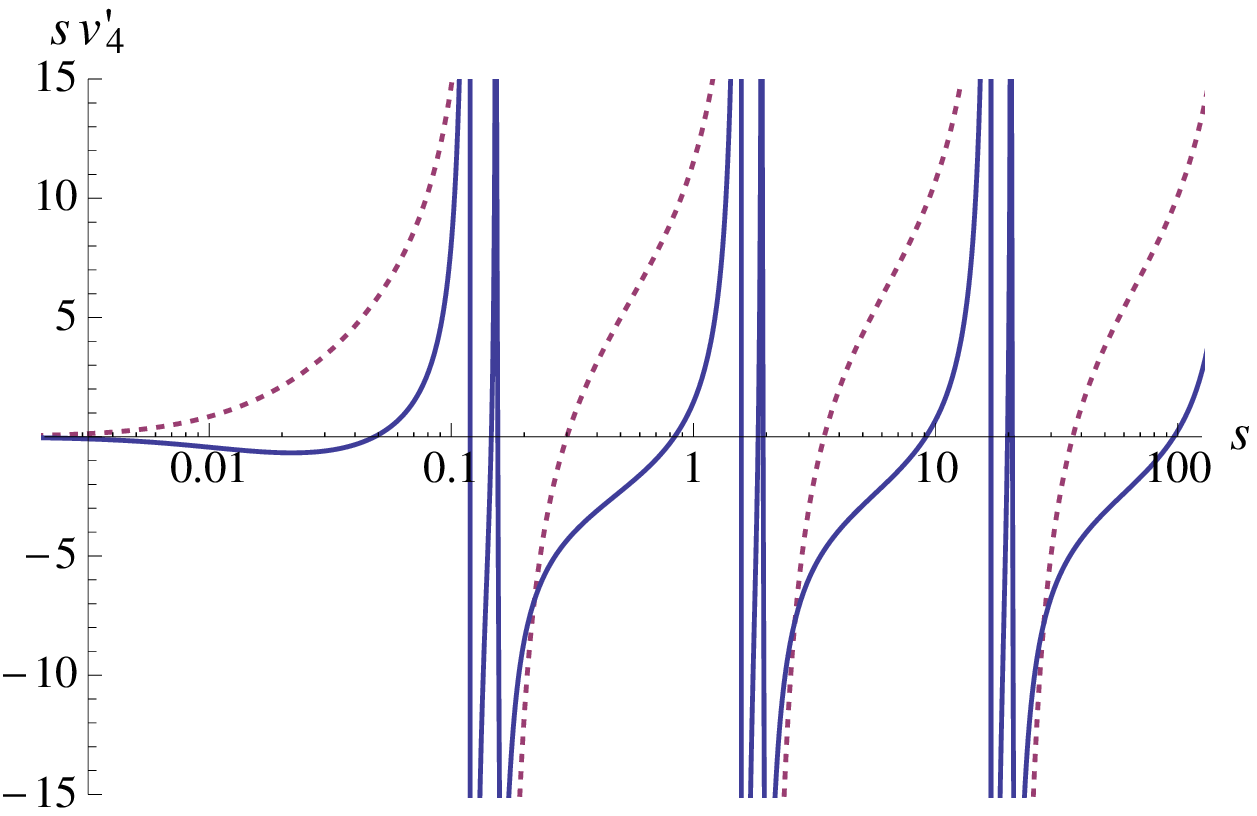}
 \caption{RG evolutions of $sv_4$, $sv_4'$ (solid curves), and $sv_3$
 (dotted curve) with initial conditions $g(0)=10$, $v_3(0)=i\,10^{-6}$,
 and $v_4(0)=v_4'(0)=0$~\cite{footnote}.  \label{fig:RG}}
\end{figure}

Similarly, the four-body couplings $v_4$ and $v_4'$ are renormalized by
diagrams depicted in Fig.~\ref{fig:4-body}.   By taking into account the
contribution from the $\phi_a$ field renormalization, the RG equations
that govern the running of $v_4$ and $v_4'$ are found to be
\begin{subequations}
 \begin{align}
  \frac{dv_4}{ds} &= -\frac{8g^4}\pi + \frac{2g^2v_3}\pi
  - \frac{2g^2v_4}\pi + \frac{2v_4^2}\pi, \\[6pt]
  \frac{dv_4'}{ds} &= -\frac{4g^4}\pi + \frac{2g^2v_3}\pi
  - \frac{2g^2v_4'}\pi + \frac{2v_4'^2}\pi,
 \end{align} 
\end{subequations}
respectively.  Numerical solutions to these RG equations are shown in
Fig.~\ref{fig:RG}, and both $sv_4$ and $sv_4'$ at large $s$ are periodic
functions of $\ln s$ with period $3\pi/4$.  We find that $v_4$
corresponding to the $\ell=0$ channel diverges only at the same points
as $v_3$.  On the other hand, $v_4'$ corresponding to the $\ell=\pm2$
channels has an additional divergence at $\ln s=3\pi n/4+\theta-0.188$
associated with every point where $v_3$ diverges.  Following the RG
study of spinless bosons in three dimensions~\cite{Schmidt:2010}, we
identify these additional divergences with four-body resonances.  We
checked that this result is independent of the choice of initial
conditions which leads to the prediction presented in
Eq.~(\ref{eq:4-body_energy}).

\begin{figure}[t]
 \includegraphics[width=\columnwidth,clip]{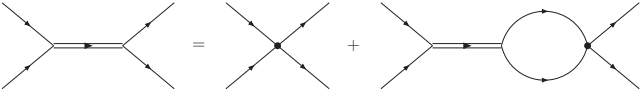}
 \caption{Two-fermion scattering with the amplitude represented by the
 double line.  \label{fig:2-fermion}}
\end{figure}

\sect{Bound state problem}
The above predictions derived by the RG analysis can be confirmed by
solving bound state problems.  We employ a model Hamiltonian
\begin{align}
 H &= \int\!\frac{d\k}{(2\pi)^2}\frac{\k^2}2\psi_\k^\+\psi_\k
 - v_0\sum_{a=\pm}\int\frac{d\k d\p d\q}{(2\pi)^6} \notag\\
 &\times \chi_a(\p)\chi_{-a}(\q)\,\psi_{\frac\k2+\p}^\+
 \psi_{\frac\k2-\p}^\+\psi_{\frac\k2-\q}\psi_{\frac\k2+\q},
\end{align}
which describes spinless fermions interacting by a separable $p$-wave
potential.  The form factor
\begin{align}
 \chi_a(\p) = p_a\,e^{-\p^2/(2\Lambda^2)}
\end{align}
is introduced to regularize the ultraviolet behavior.  By summing
diagrams depicted in Fig.~\ref{fig:2-fermion}, the two-fermion
scattering $T$ matrix with incoming $\q$ and outgoing $\p$ relative
momenta is obtained as
\begin{align}
 T(E;\p,\q) = \frac{16\pi|\p||\q|\cos(\varphi_\p-\varphi_\q)\,
 e^{-(\p^2+\q^2)/(2\Lambda^2)}}{\frac{2\pi}{v_0}-\Lambda^2
 -E\,e^{-E/\Lambda^2}\,\mathrm{E}_1(-E/\Lambda^2)},
\end{align}
where $E=\omega-\k^2/4+i0^+$ is the total energy in the center-of-mass
frame and $\mathrm{E}_1(w)\equiv\int_w^\infty dt\,e^{-t}/t$ is the
exponential integral.  Accordingly, the $p$-wave resonance is achieved
by tuning the bare two-body coupling at $v_0=2\pi/\Lambda^2$.

\begin{figure}[t]
 \includegraphics[width=\columnwidth,clip]{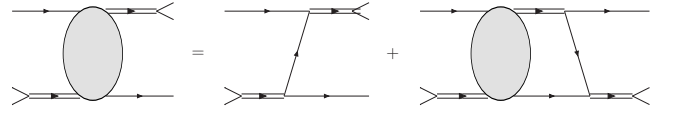}
 \caption{Three-fermion scattering with the amplitude represented by the
 blob.  \label{fig:3-fermion}}
\end{figure}

A three-fermion scattering problem can be solved in a similar way to the
corresponding problem in three
dimensions~\cite{Levinsen:2007,Jona-Lasinio:2008}.  The $T$-matrix
elements satisfy coupled integral equations depicted in
Fig.~\ref{fig:3-fermion}.  When the total energy approaches a binding
energy $E\to-\kappa^2<0$, the $T$ matrix factorizes as
$T_{ab}(E;\p,\q)\to Z_a(\p)Z_b^*(\q)/(E+\kappa^2)$, and the resulting
residue function satisfies
\begin{align}\label{eq:integral_eq}
 Z_a(\p) &= -\int\!\frac{d\q}{2\pi}\,
 \frac{(\p+2\q)_{-a}\,e^{-(5\p^2+5\q^2+8\p\cdot\q)/(8\Lambda^2)}}
 {\p^2+\q^2+\p\cdot\q+\kappa^2} \notag\\[6pt]
 &\times \frac{\sum_{b=\pm}(2\p+\q)_b\,Z_b(\q)}
 {(\frac34\q^2+\kappa^2)\,e^{(\frac34\q^2+\kappa^2)/\Lambda^2}\,
 \mathrm{E}_1[(\frac34\q^2+\kappa^2)/\Lambda^2]}.
\end{align}
It can be seen that $Z_+(\p)=e^{i(\ell-1)\varphi_\p}z_+(p)$ couples with
$Z_-(\p)=e^{i(\ell+1)\varphi_\p}z_-(p)$.  We focus on the $\ell=+1$
channel in which the super Efimov effect is expected to emerge.

We first solve the two coupled integral equations (\ref{eq:integral_eq})
analytically within the leading-logarithm approximation~\cite{Son:1999}.
We assume that the integral is dominated by the region where
$\kappa\ll q\ll\Lambda$ and split the integral into two parts,
$\kappa\ll q\ll p$ and $p\ll q\ll\Lambda$, in which a sum of $p$ and $q$
in the integrand is replaced with whichever is larger.  This
approximation simplifies Eq.~(\ref{eq:integral_eq}) to
\begin{subequations}
 \begin{align}
  \frac34z_+(p) &= -\int_\kappa^p\!\frac{dq}{q}\,\frac{z_+(q)}{\ln\Lambda/q}
  - \int_p^{\epsilon\Lambda}\!\frac{dq}{q}\,\frac{z_+(q)+z_-(q)}{\ln\Lambda/q}, \\
  \frac34z_-(p) &= -\int_\kappa^p\!\frac{dq}{q}\,\frac{z_+(q)}{\ln\Lambda/q},
 \end{align}
\end{subequations}
where $\epsilon<1$ is a positive constant.  By changing variables to
$x=\ln\ln\Lambda/p$ and $y=\ln\ln\Lambda/q$ and defining
$\lambda=\ln\ln\Lambda/\kappa$, $\eta=\ln\ln1/\epsilon$, and
$\zeta_\pm(x)=z_\pm(p)$, we obtain
\begin{subequations}\label{eq:integral_eqs}
 \begin{align}
  \frac34\zeta_+(x) &= -\int_x^\lambda\!dy\,\zeta_+(y)
  - \int_\eta^x\!dy\,[\zeta_+(y)+\zeta_-(y)], \\
  \frac34\zeta_-(x) &= -\int_x^\lambda\!dy\,\zeta_+(y).
 \end{align}
\end{subequations}
Then the differentiation of Eqs.~(\ref{eq:integral_eqs}) with respect to
$x$ results in two coupled differential equations
\begin{subequations}\label{eq:diff_eqs}
 \begin{align}
  \frac34\zeta_+'(x) &= -\zeta_-(x), \\[4pt]
  \frac34\zeta_-'(x) &= \zeta_+(x)
 \end{align}
\end{subequations}
with boundary conditions
\begin{subequations}
 \begin{align}
  \label{eq:boundary_+}
  \frac34\zeta_+(\eta) &= -\int_\eta^\lambda\!dy\,\zeta_+(y), \\[2pt]
  \frac34\zeta_-(\lambda) &= 0.
  \label{eq:boundary_-}
 \end{align}
\end{subequations}
The differential equations (\ref{eq:diff_eqs}) with the boundary
condition (\ref{eq:boundary_-}) are solved by
\begin{subequations}\label{eq:solutions}
 \begin{align}
  \zeta_+(x) &= \cos\!\left(\frac43(x-\lambda)\right), \\[4pt]
  \zeta_-(x) &= \sin\!\left(\frac43(x-\lambda)\right).
 \end{align}
\end{subequations}
The other boundary condition (\ref{eq:boundary_+}) constrains the
allowed value of $\lambda$, while we cannot determine its value within
the present approximation, because it is sensitive to the ultraviolet
physics at $q\sim\Lambda$.  However, owing to the periodicity of
solutions (\ref{eq:solutions}), if $\lambda=\theta$ is a solution, then
$\lambda=3\pi n/4+\theta$ must be all solutions, which is consistent
with our previous RG analysis.  We note that the solutions for $\ell=-1$
are obtained simply by exchanging $+\leftrightarrow-$, while the same
approximation does not yield any solution for $\ell\neq\pm1$.  The
double-logarithm scaling of solutions was also found in studying a
scattering problem of three fermions~\cite{Levinsen:2008}.

\begin{table}[t]
 \caption{Lowest six binding energies $E_3^{(n)}=-\kappa_n^2/m$ in the
 form of $\lambda_n=\ln\ln\Lambda/\kappa_n$.  \label{tab:binding}}
 \begin{ruledtabular}
  \begin{tabular}{ccccccc}
   & $n$ && $\lambda_n$ && $\lambda_n-\lambda_{n-1}$ & \\[2pt]\hline
   & 0 && \ \,0.5632 && --- & \\
   & 1 && 2.770 && 2.207 & \\
   & 2 && 5.078 && 2.308 & \\
   & 3 && 7.430 && 2.352 & \\
   & 4 && 9.785 && 2.355 & \\
   & 5 && 12.141\,\ \ && 2.356 & \\[2pt]\hline
   & $\infty$ && --- && $\ \ \ 2.35619$ & \\
  \end{tabular}
 \end{ruledtabular}
\end{table}

We also solved the two coupled integral equations (\ref{eq:integral_eq})
with $\ell=\pm1$ numerically and obtained binding energies of three
fermions shown in Table~\ref{tab:binding}.  We find that they obey the
universal doubly exponential scaling with period $3\pi/4\approx2.35619$,
which indeed confirms our prediction in Eq.~(\ref{eq:3-body_energy}).
While our prediction for $\ell=\pm2$ four-fermion resonances in
Eq.~(\ref{eq:4-body_energy}) can be tested in principle in a similar way
to the corresponding problem in the usual Efimov
effect~\cite{Hammer:2007,von-Stecher:2009}, we defer this problem to a
future study.

\sect{Conclusion}
In this Letter, we predicted the super Efimov effect in which resonantly
interacting fermions in two dimensions form an infinite tower of
$\ell=\pm1$ three-body bound states whose binding energies obey the
universal doubly exponential scaling as in Eq.~(\ref{eq:3-body_energy}).
We also argued that there are $\ell=\pm2$ four-body resonances
associated with every three-body bound state as in
Eq.~(\ref{eq:4-body_energy}).  It is possible to extend our analysis
away from the $p$-wave resonance and draw a binding energy diagram as a
function of the detuning in analogy to the usual Efimov
effect~\cite{Ferlaino:2010}.  These universal few-body states should be
taken into account in future many-body studies of the system such as the
topological quantum phase transition~\cite{Read:2002,Gurarie:2005}.  Our
super Efimov states (hopefully the lowest one) may be observed in
ultracold atom experiments with $p$-wave Feshbach
resonances~\cite{Chin:2010} in the same way as the usual Efimov states
through a measurement of three-body or four-body recombination atom
loss~\cite{Kraemer:2006,Ferlaino:2009}.

\acknowledgments
The authors thank Richard Schmidt for valuable discussions.  This work
was supported by a LANL Oppenheimer Fellowship, U.S.\ DOE Grant
No.\ DE-FG02-97ER41014, and NSF MRSEC Grant No.\ DMR-0820054.  Part of
the numerical calculations were carried out at the YITP computer
facility in Kyoto University.

\end{document}